\documentclass[aps,prc,reprint,superscriptaddress,nofootinbib]{revtex4-1}

\usepackage{graphicx,amsfonts,color,epsfig,epstopdf}
\usepackage{footnote,multirow}
\usepackage{amsmath}
\usepackage{amssymb}
\usepackage[utf8]{inputenc}

\newcommand{\SU}[1]{\ensuremath{\mathrm{SU}( #1 )}}

\newcommand{\SpR}[1]{\ensuremath{\mathrm{Sp}( #1,\mathbb{R} )}}
\newcommand{\ket}[1]{\ensuremath{\left| #1 \right\rangle}}
\newcommand{\bra}[1]{\ensuremath{\left\langle #1 \right|}}
\newcommand{\braket}[2]{\ensuremath{\left\langle #1 | #2  \right\rangle }}
\newcommand{\Nmax}{$N_\mathrm{max}$ }
\newcommand{\hw}{$\hbar\Omega$ }
\newcommand{\bt}{$\beta$ }
\newcommand{\Be}{$^8\mathrm{Be}$}
\newcommand{\Li}{$^8\mathrm{Li}$}
\newcommand{\NNLOopt}{NNLO$_\mathrm{opt}$}

\newcommand{\etal}{\emph{et al.}}

\begin{document}
\title{Impact of clustering on the \Li~\bt decay and recoil form factors
}
\author{G. H. Sargsyan}
\affiliation{Department of Physics and Astronomy, Louisiana State University, Baton Rouge, LA 70803, USA}
\author{K. D. Launey}
\affiliation{Department of Physics and Astronomy, Louisiana State University, Baton Rouge, LA 70803, USA}
\author{M. T. Burkey}
\affiliation{Department of Physics, University of Chicago, Chicago, Illinois 60637, USA}
\affiliation{Physics Division, Argonne National Laboratory, Argonne, Illinois 60439, USA}
\affiliation{Lawrence Livermore National  Laboratory, Livermore, California 94550, USA}
\author{A. T. Gallant}
\affiliation{Lawrence Livermore National  Laboratory, Livermore, California 94550, USA}
\author{N. D. Scielzo}
\affiliation{Lawrence Livermore National  Laboratory, Livermore, California 94550, USA}
\author{G. Savard}
\affiliation{Physics Division, Argonne National Laboratory, Argonne, Illinois 60439, USA}
\author{A. Mercenne}
\affiliation{Department of Physics and Astronomy, Louisiana State University, Baton Rouge, LA 70803, USA}
\affiliation{Center for Theoretical Physics, Sloane Physics Laboratory, Yale University, New Haven, Connecticut 06520, USA}
\author{T. Dytrych}
\affiliation{Department of Physics and Astronomy, Louisiana State University, Baton Rouge, LA 70803, USA}
\affiliation{ Nuclear Physics Institute of the Czech Academy of Sciences, 250 68 \v{R}e\v{z}, Czech Republic}
\author{D. Langr}
\affiliation{Department of Computer Systems, Faculty of Information Technology, Czech Technical University in Prague, Prague 16000, Czech Republic}
\author{L. Varriano}
\affiliation{Department of Physics, University of Chicago, Chicago, Illinois 60637, USA}
\affiliation{Physics Division, Argonne National Laboratory, Argonne, Illinois 60439, USA}
\author{B. Longfellow}
\affiliation{Lawrence Livermore National  Laboratory, Livermore, California 94550, USA}
\author{T. Y. Hirsh}
\affiliation{Soreq Nuclear Research Center, Yavne 81800, Israel}
\author{J. P. Draayer}
\affiliation{Department of Physics and Astronomy, Louisiana State University, Baton Rouge, LA 70803, USA}

\begin{abstract}
We place unprecedented constraints on recoil corrections in the $\beta$ decay of $^8$Li, by identifying a strong correlation between them and the $^8$Li ground state quadrupole moment in large-scale \textit{ab initio} calculations.
The results are essential for improving the sensitivity of
high-precision experiments that probe 
the weak interaction theory and test physics beyond the Standard Model (BSM). In addition, our calculations predict a $2^+$ state of the $\alpha+\alpha$ system that is energetically accessible to $\beta$ decay but 
has not been observed in the experimental $^8$Be energy spectrum, and has an important effect on the recoil corrections and 
$\beta$ decay
for the $A=8$ systems. This state and an associated $0^+$ state are notoriously difficult to model due to their cluster structure and collective correlations, but become feasible for calculations in the \textit{ab initio} symmetry-adapted no-core shell-model framework.

\end{abstract}

\maketitle

\noindent
\emph{Introduction.--} The left-handed vector minus axial-vector (V$-$A) structure of the weak interaction was postulated in late 1950's and early 1960's \cite{feynman1958theory, sudarshan1958chirality} guided in large part by a series of $\beta$-decay experiments  \cite{Wu1957, Herrmannsfeldt1957,johnson1963precision}, and later was incorporated in the Standard Model of particle physics. However, in its most general form, the weak interaction can also have scalar, tensor, and pseudoscalar  terms as well as right-handed currents. 

Today, $\beta$-decay experiments continue to pursue increasingly sensitive searches for additional contributions to the weak interaction.
Various experiments \cite{johnson1963precision,FlechardVLM2011,LiSSB2013CT, sternberg2015limit} have constrained the tensor part of the interaction, although the limits are less stringent compared to the other non-standard-model terms \cite{severijns2006tests,wauters2014limits}. While these experiments have achieved remarkable precision, further improvements require confronting the systematic uncertainties that stem from higher-order corrections (referred to as recoil-order terms) in nuclear \bt decay. These terms are inherently small compared to the allowed \bt decay terms; however, current experiments have reached a precision where even subtle distortions matter. Measurements of recoil-order terms are also interesting in their own right as they can test additional symmetries of the Standard Model, such as the existence of second-class currents \cite{Grenacs1985, DeBraeckeleer1995, sumikama2008search} and the accuracy of the conserved vector current (CVC) hypothesis \cite{Holstein1974,Tribble1975,DeBraeckeleer1995,sumikama2011test}. 
\begin{figure}[ht]
    \centering
    \includegraphics[width=0.40\textwidth]{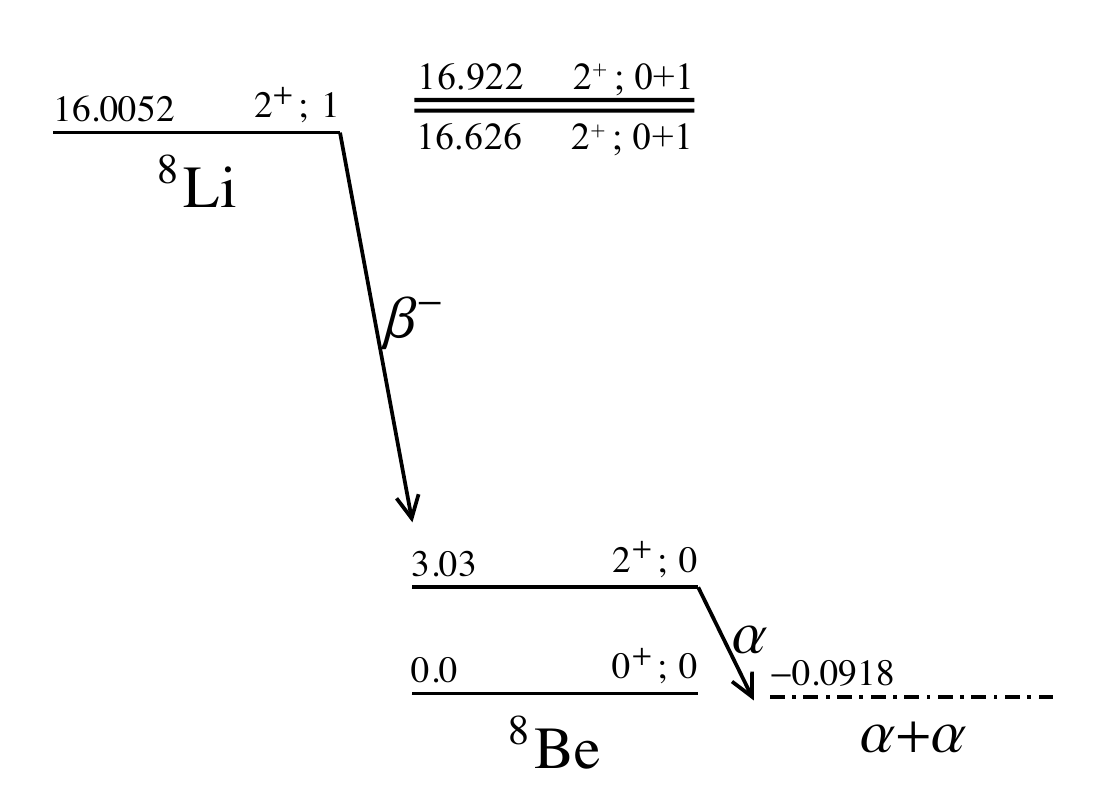}
    \caption{Decay scheme for \bt decay of \Li{} $\mathrm{g.s.}$ to the broad low-lying $2^+$ state in \Be.
    Energies, in MeV, are relative to \Be~$\mathrm{g.s.}$ A small $\beta$-decay branching is observed to the doublet 2$^+$ states due to their resonant nature.
    }
    \label{fig:IAS}
\end{figure}

The $\beta$ decay of $^8$Li to \Be, which subsequently breaks up into two $\alpha$ particles, has long been recognized as an excellent testing ground to search for new physics~\cite{Barnes1958,Holstein1974,Grenacs1985} due to the high decay energy and the ease of detecting the $\beta$ and two $\alpha$ particles. Recently, by taking advantage of ion-trapping techniques, high-precision measurements of $\beta$-$\bar{\nu}$-$\alpha$ correlations~\cite{ScielzoLSSB2012beta,LiSSB2013CT,sternberg2015limit} have been performed that set the most stringent limit on a tensor contribution to date \cite{Burkey2021prl}. However, in this type of experiment, one of the largest uncertainties comes from the several different recoil-order corrections that contribute to the decay.
A number of other experiments have taken advantage of the presence of certain recoil-order terms in the \bt decay of \Li~(Fig. \ref{fig:IAS}) to perform CVC tests by studying $\beta$-$\alpha$ angular correlations \cite{Tribble1975,McKeownGG1980} and $\beta$-spin alignment correlations \cite{sumikama2008search,sumikama2011test}. 
In addition, these terms have been deduced from $\gamma$ decays of the doublet 2$^+$ states near 17 MeV in the \Be{} spectrum, which contain the isobaric analogue of the \Li~ground state ($\mathrm{g.s.}$) \cite{DeBraeckeleer1995}. Due to their small size, and the fact that there are several terms that contribute to decay observables, most of the experimentally extracted recoil-order terms have large uncertainties.

In this Letter, we report the first \textit{ab initio} calculations of recoil-order terms in the \bt decay of \Li. These calculations achieve highly reduced uncertainties compared to the experimentally extracted values of  \cite{sumikama2011test}. They help decrease the systematic uncertainties on the tensor-current estimates in the weak interaction reported in \cite{Burkey2021prl}, and are of interest to experimental tests of the CVC hypothesis   \cite{DeBraeckeleer1995}.
We also provide evidence that the $\beta$-transition strength of the \Li~decay
is affected by a disputed low-lying  2$^+$ state (sometimes referred to as an ``intruder" state) below 16 MeV in the \Be{} spectrum. Our calculations in unprecedentedly large model spaces support the existence of low-lying states with a large overlap with the 
$\alpha+\alpha$ $s$- and $d$-waves. Indeed, a very broad 2$^+$ state along with a lower 0$^+$ were initially proposed by Barker from the R-matrix analysis of $\alpha+\alpha$ scattering and the \bt decays of \Li~and $^8$B \cite{Barker1968, Barker1969, Barker1989}. Even though such states have not been directly observed experimentally,
some earlier theoretical studies have predicted them in the low-lying spectrum of \Be~\cite{Caurier2001, Maris2013, Rodkin2020}. Furthermore, there has been a recent experimental indication in favor of intruder states below 16 MeV \cite{Munch2018PLB}.  

\noindent
 \textit{SA-NCSM framework.--} 
 For this work, we employ the \textit{ab initio} symmetry-adapted no-core shell model (SA-NCSM) \cite{LauneyDD16,DytrychLDRWRBB20,LauneyMD_ARNPS21}. The use of chiral effective-field-theory interactions \cite{BedaqueVKolck02,EpelbaumNGKMW02,EntemM03,Epelbaum06} 
 enables nuclear calculations 
informed by elementary particle physics, while the symmetry-adapted (SA) basis allows us to achieve ultra-large model spaces imperative for the description of challenging features in the \Be ~states, such as clustering and collectivity. 
It uses a harmonic oscillator (HO) basis with frequency \hw~and a model space with an \Nmax cutoff, which is the 
 maximum  total HO excitation quanta above the lowest HO configuration for a given nucleus. 
 These parameters 
 are related to infrared (IR) and ultraviolet (UV) cutoffs \cite{WendtFPS15}, which can be understood as the effective size of the model space in which the nucleus resides, and its grid resolution, respectively.  The calculations become independent of \hw at \Nmax$\rightarrow\infty$, providing a parameter-free \textit{ab initio} prediction. 
 The SA-NCSM results exactly reproduce those of the NCSM \cite{NavratilVB00,NavratilVB00b} for the same nuclear interaction. However, by utilizing the emergent symplectic \SpR{3}-symmetry in nuclei \cite{DytrychLDRWRBB20}, the SA-NCSM can expand the model space by a physically relevant subspace, which is only a fraction of the complete NCSM space, thereby including localized-$\alpha$ 
 degrees of freedom 
 within the interaction effective range  \cite{DreyfussLESBDD2020}.
 \begin{figure*}[ht]
   \includegraphics[width=0.99\textwidth]{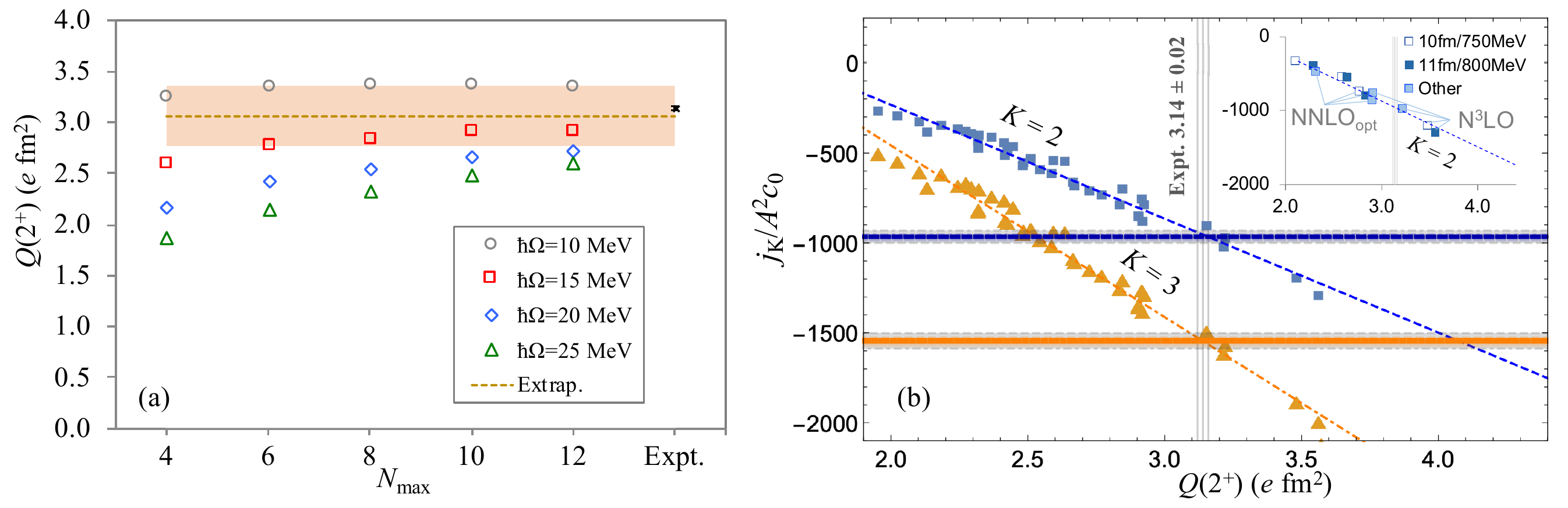}
    \caption{(a) \textit{Ab initio} \Li~g.s. quadrupole moment $Q(2^+)$ compared to experiment \cite{BorremansBBG2005Q2} (denoted as ``Expt.''). Calculations 
    use the NNLO$_\mathrm{opt}$ NN for different model-space sizes and resolutions (open symbols), along with the infinite-size extrapolated value (dashed line) and the corresponding many-body uncertainty (shaded area). 
    (b) Calculated $j_2/A^2c_0$ and $j_3/A^2c_0$ (squares and triangles, respectively) and their predicted values (upper and lower horizontal 
    lines, respectively) for the $^8\mathrm{Li}\; \beta$ decay to  $2_1^+$ in \Be{} vs. the calculated \Li~$Q(2^+$).  The first (second) set of the uncertainties in Eq. (\ref{pred_jK}) is shown as the line thickness (gray bands). Calculations use the NNLO$_\mathrm{opt}$, NNLO$_\mathrm{sat}$ and ${\rm N}^{3}{\rm LO}$ chiral potentials, and the JISP16 NN, in \Nmax=6-12 model spaces. 
   Inset: subset of calculated $j_2/A^2c_0$ vs. $Q(2^+$) for selected IR (in fm)/UV (in MeV) many-body cutoffs across all interactions, and for two interactions across several cutoffs. 
    }
    \label{fig:jk_vs_Q}
\end{figure*}

 We adopt various chiral 
potentials without renormalization in nuclear medium: 
 N$^3$LO-EM \cite{EntemM03},  NNLO$_{\rm opt}$ \cite{Ekstrom13}, as well as NNLO$_\mathrm{sat}$ \cite{NNLOsat2015} with the three-nucleon (3N) forces, hierarchically smaller than their nucleon-nucleon  (NN) forces, added as averages \cite{LauneyMD_ARNPS21}. For comparison, we present results with the soft JISP16 phase-equivalent NN interaction \cite{Shirokov2010nn}. 
  We use \hw{}=15-25 MeV for  N$^3$LO-EM, NNLO$_{\rm opt}$, and JISP16, and \hw{}=16 and 22 MeV for  NNLO$_\mathrm{sat}$, unless otherwise stated.
 The NNLO$_{\rm opt}$ is used without 3N forces, which have been shown to contribute minimally to the 3- and 4-nucleon binding energy \cite{Ekstrom13}. Furthermore, the NNLO$_{\rm opt}$ NN potential has been found to reproduce various observables, including the $^4$He electric dipole polarizability \cite{BakerLBND20}; the challenging analyzing power for elastic proton scattering on $^4$He, $^{12}$C, and $^{16}$O \cite{BurrowsEWLMNP19}; along with  B(E2) transition strengths  for $^{21}$Mg and $^{21}$F  \cite{Ruotsalainen19} in the SA-NCSM without effective charges.

For the purposes of this study,
the quadrupole moment of the \Li~$\mathrm{g.s.}$, $Q(2_\mathrm{g.s.}^+)$, for which SA-NCSM calculations with the \NNLOopt{} NN are extrapolated to an infinite model-space size, 
is shown to reproduce the experimental value within the many-body model uncertainties (Fig. \ref{fig:jk_vs_Q}a). The result is in close agreement with the extrapolated value of \cite{McCrackenNMQH2021} that uses renormalized NN+3N chiral potentials. The model uncertainties are based on variations in the model-space size and resolution, and extrapolations use the Shanks method \cite{DytrychLDRWRBB20}.

\noindent
\emph{Recoil-order corrections.--} The recoil-order form factors are generally neglected in $\beta$-decay theory since they are of the order of $q/m_N$ or higher, where $q$ is the momentum transfer (typically several MeV/c) and $m_N$ is the nucleon mass \cite{Holstein1974}. Thus, for most \bt decays, the recoil effects are typically less than a percent of the dominant Fermi and Gamow-Teller (GT) contributions (for an example see \cite{Glick-MagidFGGG2021}). However, for measurements of sufficiently high precision, these terms must be included in the analysis especially when the leading contributions are suppressed or the recoil-order terms are unusually large. 
These recoil-order form factors include, the second forbidden axial vectors ($j_{2}$ and $j_{3}$), induced tensor ($d$), and weak magnetism ($b$), 
and along with the  GT ($c_0$) are given in the impulse approximation (IA) as:
\begin{widetext}
\begin{eqnarray}
 c_0(q^2) &=& (-)^{(J'-J)} \frac{g_A(q^2)}{\sqrt{2J+1}} \bra{J'} |  \sum_{i=1}^A \tau^{\pm}_i  \sigma_i  | \ket{J} = (-)^{(J'-J)} \frac{g_A(q^2)}{\sqrt{2J+1}} M_{GT}, \nonumber \\
j_K(q^2) &=& - (-)^{(J'-J)} \frac{2}{3} \frac{g_A(q^2)}{\sqrt{2J+1}} \frac{(Am_N c^2)^2 }{(\hbar c)^2}  \bra{J'} |  \sum_{i=1}^A \tau^{\pm}_i  [Q_i \times \sigma_i]^K | \ket{J}, {\rm with\,} K=2,3, \nonumber \\
d(q^2) &=& (-)^{(J'-J)} A\frac{g_A(q^2)}{\sqrt{2J+1}} \bra{J'} |  \sum_{i=1}^A \tau^{\pm}_i  \sqrt{2}[L_i \times \sigma_i ]^1 | \ket{J}, \nonumber \\
b(q^2) &=& A\frac{(-)^{(J'-J)} }{\sqrt{2J+1}} \Big[ g_M (q^2)\bra{J'} |  \sum_{i=1}^A \tau^{\pm}_i  \sigma_i  | \ket{J}   +g_V(q^2) \bra{J'}|  \sum_{i=1}^A \tau^{\pm}_i L_i  | \ket{J}\Big],
\label{eq:recoil}
\end{eqnarray}
\end{widetext}
where 
$g_V(0)=1$, $g_A(0)\approx 1.27$ and  $g_M(0) \approx 4.70$ are the vector, axial and weak magnetism coupling constants, $A$ is the mass number and $J (J')$ is the total angular momentum of the initial (final) nucleus. The $\tau_i$/2, $\sigma_i/2$, $Q_i=\sqrt{16\pi/5}r_i^2Y_{2\mu}(\hat r_i)$, and $L_i$ are the isospin,  intrinsic spin, quadrupole moment and angular momentum operators, respectively, of the $i^\mathrm{th}$ particle. 
$M_{GT}$ is the conventional GT matrix element. The matrix elements in Eq. (\ref{eq:recoil}) are computed translationally invariant in the SA-NCSM. 
These recoil-order form factors, usually reported as the
ratios $j_{2,3}/A^2c_0$, $d/Ac_0$, and $b/Ac_0$, enter into the expression of the $\beta$-decay rate for nuclei undergoing delayed $\alpha$-particle emission \cite{Holstein1974, sternberg2015limit, Burkey:2019ant, Burkey2021prl}. 

Remarkably, we identify a strong correlation between $j_{2,3}/A^2c_0$ and the \Li~$\mathrm{g.s.}$ quadrupole moment based on calculations across several  interactions, \Nmax{} and \hw parameters (Fig. \ref{fig:jk_vs_Q}b, using \Nmax= 6 to 12 for NNLO$_\mathrm{opt}$, to 10 for N3LO-EM and JISP16, and to 8 for NNLO$_\mathrm{sat}$). 
As can be seen in the Fig. 2b inset, the linear dependence is observed regardless of any errors that may arise from the many-body truncation and from the higher-order effects (e.g., \cite{FilinMBEKR2021,MarisEFG2021}) associated with various interactions. An identical spread is found for $j_3/A^2 c_0$ due to the strong correlation between $j_2$ and $j_3$ (see Supplemental Material \cite{suppl}). \nocite{suhonen2007nucleons, Navratil04,CastanosDL88,MustonenGAB2018}
A linear regression along with the combination of the correlation to $Q(2_{\rm g.s.}^+)$ and its experimental value of 3.14(2) $e \,\rm{fm}^2$ \cite{BorremansBBG2005Q2} lead to reduced uncertainties on our predictions:
\begin{equation}
  \frac{j_2}{A^2c_0}=-966 \pm 13 \pm 33, \quad \frac{j_3}{A^2c_0}=-1546 \pm 19 \pm 40,
    \label{pred_jK}
\end{equation}
\noindent Here, the first set of uncertainties uses the quadrupole moment experimental uncertainties given the linear regression slope, and the second set arises from the regression uncertainty using Student's $t$-distribution and a $99\%$ confidence level.
This correlation is important, as we can reduce the problem of calculating a matrix element that depends on cluster physics in \Be{} to a bound state observable in $^8$Li. 

Most significantly, with the values in Eq. (\ref{pred_jK}), the uncertainty from the recoil-order corrections on the tensor current contribution to the weak interaction presented in \cite{Burkey:2019ant} is reduced by over $50\%$ \cite{Burkey2021prl}. The recoil-order terms, including the $b$ and $d$ terms, for the lowest four SA-NCSM $2^+$ states, are summarized in Table \ref{tb:recoil}.
The $d/Ac_0$ prediction for $2_1^+$  is based on a  correlation similar to the one for $j_{2,3}/A^2c_0$
(see Supplemental Material \cite{suppl}). These predictions can be used in future experiments to constrain BSM tensor currents, while these $b$ weak magnetism
predictions are of interest to experiments that test the CVC hypothesis and $d$ is of importance to determining the existence of second-class currents
 \cite{DeBraeckeleer1995}.   
\begin{table}[h!]
\caption{The recoil-order terms from SA-NCSM. Results for the 2$^+_1$ $j_{2,3}/A^2c_0$  and $d/Ac_0$ are based on the correlation to $Q(2_{\rm g.s.}^+)$; all other calculations use \NNLOopt{} and have error bars from variations in \hw{} by 5 MeV and in model-space sizes up to \Nmax=16 (12) for $j_{2,3}/A^2c_0$ ($d/Ac_0$ and $b/Ac_0$).
}
\label{tb:recoil}

\begin{tabular}{lllll}
\hline \hline
\multicolumn{1}{c}{} & \multicolumn{1}{c}{$j_2/A^2c_0$} & \multicolumn{1}{c}{$j_3/A^2c_0$} & \multicolumn{1}{c}{$d/Ac_0$} & \multicolumn{1}{c}{$b/Ac_0$} \\ \hline
$2_1^+$             &  $-966     \pm 36$              & $-1546 \pm 44$                  & $10.0 \pm 1.0$                                     & $6.0 \pm 0.4 $                       \\
$2^+_2 \mathrm{(new)}$           & $-10  \pm 10$                 & $-80  \pm 30$                & $-0.5  \pm 0.5$                    & $3.7  \pm 0.4  $                      \\
$2^+_3 \rm{(doublet\, 1)}$            & $12  \pm 5$                  & $-60  \pm 15  $                & $0.3  \pm 0.2  $                   & $3.8  \pm 0.2$                         \\
$2^+_4 \rm{(doublet\, 2)}$       & $11  \pm 3$                   & $-65 \pm 11$                 & $0.2 \pm 0.2$                 & $3.8 \pm 0.2$                     \\
\hline \hline
\end{tabular}
\end{table}

\noindent
\textit{New final state for \bt decays to} \Be{}.-- 
The experimentally deduced values presented in \cite{sumikama2011test}, $j_2/A^2c_0=-490\pm70$, $j_3/A^2c_0=-980\pm280$, $d/Ac_0=5.5 \pm 1.7$ and $b/Ac_0=7.5 \pm 0.2$, are comparable but different from our predicted values.
These experimental results were obtained through a global
fit to $\beta$-spin alignment \cite{sumikama2011test} and $\beta$-$\alpha$ angular correlation data \cite{McKeownGG1980} from \Li~and $^8$B \bt decays. Due to the small size of higher-order effects and relatively large statistical uncertainties, the $j_{2,3}/A^2c_0$ and $d/Ac_0$ were assumed in \cite{sumikama2011test} to be independent of the \Be~excitation energy. Thus, the results were averaged over the entire \bt decay spectrum. In contrast,
the SA-NCSM  wavefunctions are for individual states, hence, the predictions in Eq. (\ref{pred_jK}) are for the lowest 2$^+$ state only, 
which is the dominant transition for the \Li{} \bt decay. 
The SA-NCSM  reveals large differences between the recoil-order terms to the lowest 2$_1^+$ and higher-lying states, the most notable being for the $j_K/A^2c_0$ terms where the values differ by almost two orders of magnitude (see Table \ref{tb:recoil}). 
Hence, the angular-correlation experiment in \cite{Burkey2021prl} minimizes the sensitivity to the higher-lying states by restricting their analysis to decays centered on the broad $2_1^+$ state. 

\begin{figure*}[ht]
    \centering
   \includegraphics[width=0.99\textwidth]{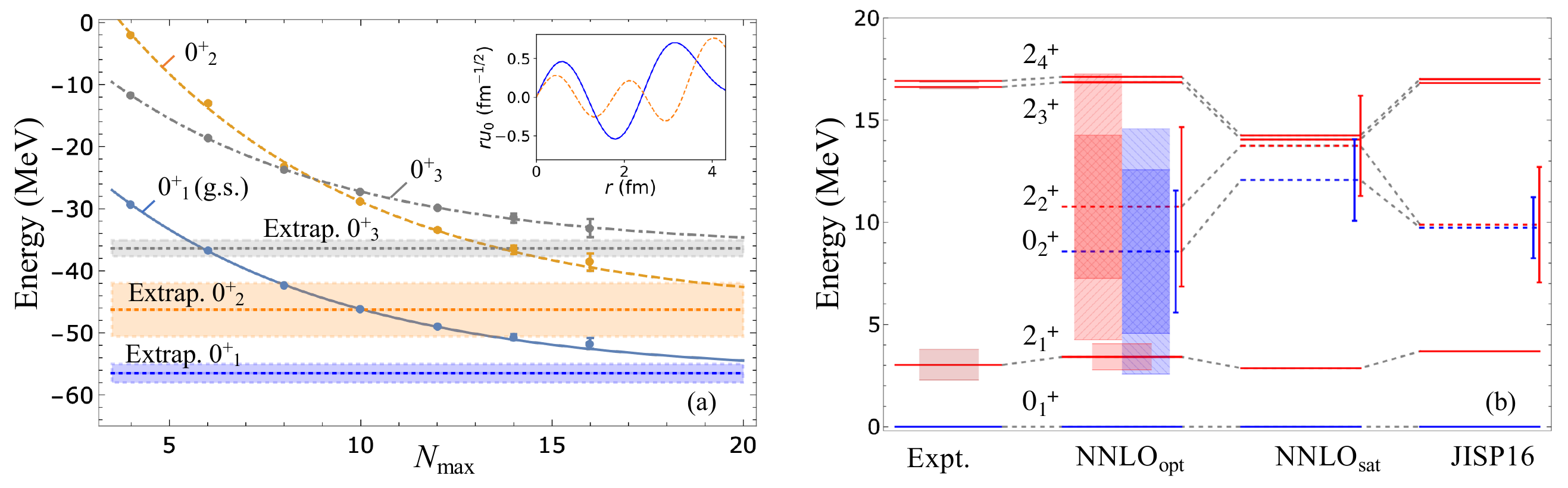}
    \caption{(a) Calculated \Be{} low-lying $0^+$ state energies illustrated for the NNLO$_\mathrm{opt}$ chiral potential (\hw=15 MeV) vs. the model-space size, together with the extrapolation values (dotted lines) and uncertainties (bands). Extrapolations use complete model spaces up to \Nmax=12 and do not include the \Nmax=14 and 16 SA selected model spaces shown with uncertainties determined by the selection. The measured $0_{\rm g.s.}^+$ energy is $-56.5$ MeV \cite{HuangWKAN2021ame}. 
    Inset: $\alpha+\alpha$ $s$-wave for $0^+_{\rm g.s.}$ (blue solid) and $0^+_2$ (orange dashed).
    (b) \textit{Ab initio} low-lying states from extrapolations for
\Be{}, compared to experiment (Expt.). The extrapolation uncertainties (error bars) for the $0^+_2$ and $2^+_2$  states (dashed levels) are based on variations in the model-space size and selection. For \NNLOopt{}, $\alpha$ width estimates (shaded areas) for the lowest two $0^+$ and $2^+$ states are shown with uncertainties (lighter shades) determined from the energy extrapolation uncertainties;
the small $0_{\rm g.s.}^+$ width (not shown) is estimated to be 5.7 eV, compared to 5.57 eV \cite{TILLEY2004155}. 
    }
    \label{fig:JJ0}
\end{figure*}

Importantly, the SA-NCSM  indicates the existence of another $2^+$ state below 16 MeV --  accessible to the \Li{} or $^8$B \bt decays through allowed transitions -- and a corresponding lower $0^+$ state that largely overlaps with the $\alpha$+$\alpha$ system (Fig. \ref{fig:JJ0}, see also Supplemental Material \cite{suppl}). 
In the SA-NCSM, these 
states 
quickly decrease in energy as the model space increases (Fig. \ref{fig:JJ0}a) regardless of the realistic  interaction used, similar to the Hoyle-state rotational band in $^{12}$C \cite{DreyfussLTDB13}. 
The extrapolations are performed using the three-parameter exponential formula from \cite{MarisVS2009}.
Notably, $0_3^+$ converges to $20.1\pm1.5$ MeV and has a structure similar to the doublet 2$^+$ states and isospin $T=1$. This state is not seen in the currently-available experimental spectrum and it is likely to be the isobaric analogue of the low-lying 0$^+$ state in \Li{} predicted by recent \textit{ab initio} calculations \cite{NavratilRQ2010,McCrackenNMQH2021}. 

The calculated low-lying states in \Be{} are in good agreement with experiment (Fig. \ref{fig:JJ0}b).
The NNLO$_\mathrm{sat}$ results include the average 3N contribution determined for a given isospin (for \Be, the contribution to the binding energy in \Nmax=12 is 1.51 MeV, resulting in a total extrapolated binding energy of 56.8 MeV). 
The extrapolations  determine the energies of $0_2^+$ and $2_2^+$  between 5 and 15 MeV above the $\mathrm{g.s.}$, corroborating earlier estimates \cite{Barker1968, Barker1969,Caurier2001}. 

For \NNLOopt{} NN and the case of the fastest energy convergence of the $0^+_2$ and $2^+_2$ states
(\hw=15 MeV),
we estimate $\alpha$ widths 
(Fig. \ref{fig:JJ0}b), by projecting the \Nmax=16 SA-NCSM wavefunctions onto $\alpha+\alpha$ cluster states and considering the exact continuum Coulomb wavefunctions outside of the interaction effective range, following the procedure of \cite{DreyfussLESBDD2020}. For this, the \Be~and $^4\mathrm{He}$ states are expressed in the \SpR{3} basis, associated with intrinsic shapes \cite{DytrychLDRWRBB20}. For \Be, we consider three dominant prolate shapes with contributions of 75\%, 4\%, and 3\% to  $0_{\rm g.s.}^+$ (totaling 82\%), and 46\%, 15\%, and 11\%  to $0_2^+$ (totaling 72\%), and similarly for the $2^+$ states (see Supplemental Material \cite{suppl}). These shapes extend to 18 HO shells and start at the most deformed configurations among those in the valence shell: 2\hw~and 4\hw~excitations. Except for the $0_{\rm g.s.}^+$ width that uses  the experimental threshold of $-92$ keV relative to the \Be~g.s., all the widths use the $\alpha$+$\alpha$ threshold estimated  at $-104$ keV  from the SA-NCSM  extrapolations of the $^4$He and $^8$Be binding energies with \NNLOopt{}. 
These widths are in good agreement with experimentally deduced values \cite{TILLEY2004155} and earlier theoretical studies \cite{KravvarisV2017, KravvarisQHN2020, ElhatisariLRE15}.

Intruder 0$^+$ and 2$^+$ states in the low-lying spectrum of \Be{} were proposed in the late 1960's by Barker from concurrent R-matrix fits to scattering, reaction, and decay data associated with the \Be{} nucleus \cite{Barker1968,Barker1969}.  
The inclusion of an intruder 2$^+$ state below 16 MeV in the R-matrix fits of \bt decays 
in \cite{Warburton1986} results in a decrease of the extracted $M_{\rm GT}$ for a decay to $2_1^+$ by almost 1.5 times,
which yields a closer agreement with the SA-NCSM $M_{\rm GT}$ (see Table \ref{tb:GT}; the calculated $M_{\rm GT}$ are not used in the experimental analysis of \cite{Burkey2021prl}). Note that, depending on the interaction, two-body axial currents may significantly affect $M_{\rm GT}$ \cite{KingAPP2020}, however, here we are interested only in the IA part. Due to the large uncertainty on the $2_2^+$ state in the calculations, we provide only the lower limits on the $\mathrm{log}(ft)$ based on the convergence pattern.  
The energies from Barker's R-matrix fits for the intruder 0$^+$ and 2$^+$ states are  $\sim 6$ MeV and 9 MeV, respectively, with $\alpha$ widths $>7$ MeV. 
These excitation energies agree with the SA-NCSM extrapolated results given the error bars (Fig. \ref{fig:JJ0}b), as well as with the predicted widths. The strong excitation-energy dependence of the recoil-order terms due to the presence of $2_2^+$ has a small effect on the weak tensor currents constraints in the low excitation-energy range (see 
systematic uncertainty in Table I in \cite{Burkey2021prl}), but is imperative to consider in analyses over the entire $\beta$ decay spectrum. 
\begin{table}[ht]
\caption{\textit{Ab initio} $M_{\rm GT}$, $c_0$ and log$(ft)$ in IA, compared to the experimentally deduced values. 
Ref. \cite{Warburton1986} includes evaluations both with an intruder $2^+$ state (denoted by *) around 8 MeV similar to Ref. \cite{Barker1989}, and without it. 
}
\label{tb:GT}

\begin{tabular}{lllll}
\hline \hline
                     & \multicolumn{3}{c}{$2_1^+$}                                                     & \multicolumn{1}{c}{$2^+_2$}                                                \\ \hline
\multicolumn{1}{c}{} & \multicolumn{1}{c}{$|M_{\rm GT}|$} & \multicolumn{1}{c}{$|c_0|$} & \multicolumn{1}{c}{log$(ft)$} & \multicolumn{1}{c}{log$(ft)$} \\ \hline
\NNLOopt             & $0.16(1)$                & $0.09(1)$                   & $5.90$                                                  & $\quad >5.06$                         \\
$\mathrm{NNLO}_\mathrm{sat}$              & $0.21(3)$                   & $0.12(2)$                   & $5.64 $                     & $\quad >5.05$                         \\
$\mathrm{JISP16} $              & $0.23(4)$                 & $0.13(2)$                   & $5.54$                     & $\quad >4.28$                         \\
Expt., Ref. \cite{Barker1989}        & $0.190 $                  & $0.108 $                 & $5.72$                  & $\quad 5.27$                      \\
Expt., Ref. \cite{Warburton1986}$^{*}$   & $0.204$                  & $0.116   $               & $5.66$                      & $\quad 5.2$             \\ 
Expt., Ref. \cite{Warburton1986}   & $0.284$                  & $0.163$                  & $5.37$                      & $\quad -$                         \\
\hline \hline
\end{tabular}
\end{table}

\noindent
\textit{Summary}.-- The \textit{ab initio} SA-NCSM has determined the size of the recoil-order form factors in the \bt decay of $^8$Li. It has shown that states of the $\alpha+\alpha$ system not included  in  the  evaluated $^8$Be energy spectrum have an important effect on all $j_{2,3}/A^2c_0$, $b/Ac_0$ and $d/Ac_0$ terms, and can explain the $M_{\rm GT}$ discrepancy in the $A=8$ systems.  The outcomes reduce -- by over $50\%$ -- the uncertainty on these recoil-order corrections,
and help improve the sensitivity of high-precision $\beta$-decay experiments that probe the V$-$A structure of the weak interaction \cite{Burkey2021prl}. 
Furthermore, our predicted  $b/Ac_0$ and $d/Ac_0$ values are important for other investigations of the Standard Model symmetries, such as the CVC hypothesis and the existence of second-class currents. 

\vspace{6pt}
\emph{Acknowledgments.-} We thank Scott Marley and Konstantinos Kravvaris for useful discussions. This work was supported in part by the U.S. National Science Foundation (PHY-1913728), SURA, Czech Science Foundation (22-14497S), and the Czech Ministry of Education, Youth and Sports under Grant No. CZ.02.1.01/0.0/0.0/16\_019/0000765. It benefited from high performance computational resources provided by LSU (www.hpc.lsu.edu), the National Energy Research Scientific Computing Center (NERSC), a U.S. Department of Energy Office of Science User Facility operated under Contract No. DE-AC02-05CH11231, as well as the Frontera computing project at the Texas Advanced Computing Center, made possible by National Science Foundation award OAC-1818253. L. Varriano was supported by a National Science Foundation Graduate Research Fellowship under Grant No. DGE-1746045.
This work was performed under the auspices of the U.S. Department of Energy by Lawrence Livermore National Laboratory under Contract DE-AC52-07NA27344.

\newpage
\section{Supplemental Material}

\subsection{Correlation of the  induced tensor term $d/Ac_0$ with $Q(2_{\rm g.s.}^+)$ of \Li}

For completeness, we present results for the correlation of the  induced tensor term $d/Ac_0$, defined in Eq. (\ref{eq:recoil}),  with the quadrupole moment of the  \Li~$\mathrm{g.s.}$ $Q(2_{\rm g.s.}^+)$. We note that all of the reduced matrix elements in Eq. (\ref{eq:recoil}) are with respect to the total angular momentum but not isospin. The additional factor, $(-)^{(J'-J)}/\sqrt{2J+1}$, compared to Ref. \cite{Holstein1974} arises from the difference in the convention of Wigner-Eckart theorem. The convention we use can be found in Ref. \cite{suhonen2007nucleons}. 

\begin{figure}[ht]
    \centering
   \includegraphics[width=0.45\textwidth]{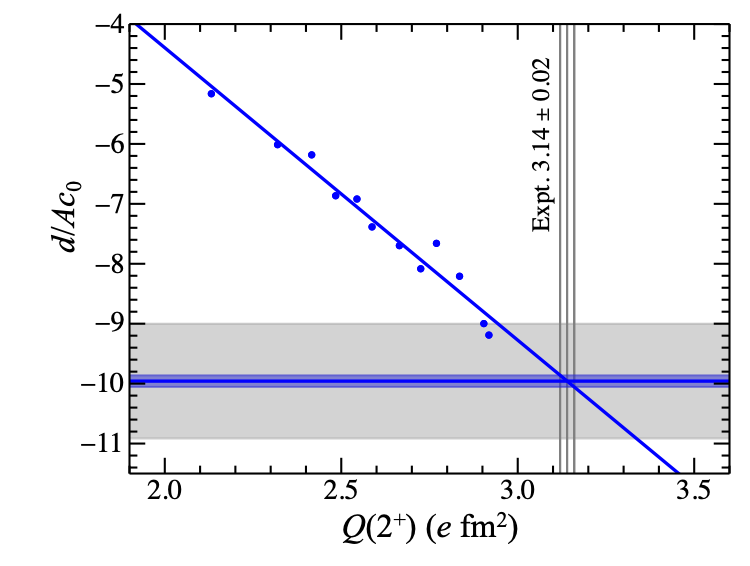}
    \caption{Calculated $d/Ac_0$ for the \Li~$\beta$ decay to the lowest $2^+$ state in \Be{} vs. the calculated \Li~ground state quadrupole moment Q(2$^+$). Calculations use the NNLO$_\mathrm{opt}$ chiral potential in model spaces \Nmax=6 to 12. The horizontal 
    line corresponds to the $d/Ac_0$ value predicted from the linear regression using the experimental \Li~quadrupole moment  
    \cite{BorremansBBG2005Q2}. The blue band indicates the uncertainty from experiment.
    The total uncertainty, which also includes the linear regression slope uncertainty, is shown by the gray band.}
    \label{fig:d_vs_Q}
\end{figure}

Similar to the $j_K/A^2c_0$ terms, the SA-NCSM calculations for $d/Ac_0$  with \NNLOopt~ for the lowest $2^+$ state in \Be~exhibit a strong correlation with the \Li~$\mathrm{g.s.}$ quadrupole moment calculations. The $d/Ac_0$ value in Table \ref{tb:recoil} is the prediction using a linear regression 
and the experimental value for the \Li~$Q(2_{\rm g.s.}^+)$ (Fig. \ref{fig:d_vs_Q}). The calculations are performed in model spaces with up to \Nmax=12 and for \hw=15, 20 and 25 MeV. The quoted value of $10.0 \pm 1.0$ has a comparatively large regression uncertainty due to the comparatively small number of calculations with a single chiral potential. This uncertainty  is calculated using the slope uncertainty multiplied by the $Q(2_{\rm g.s.}^+)$ experimental  value. 
Further improvement of the $d/Ac_0$ uncertainty is underway: with the present study reducing the largest uncertainties hitherto, namely, for the $j_2/A^2c_0$  and $j_3/A^2c_0$ terms, the $d/Ac_0$ uncertainty  becomes now important for further investigations of tensor currents and other physics beyond the Standard Model \cite{Burkey2021prl},

\subsection{Correlation of $j_2/A^2c_0$ with $j_3/A^2c_0$}

\begin{figure}[ht]
    \centering
   \includegraphics[width=0.48\textwidth]{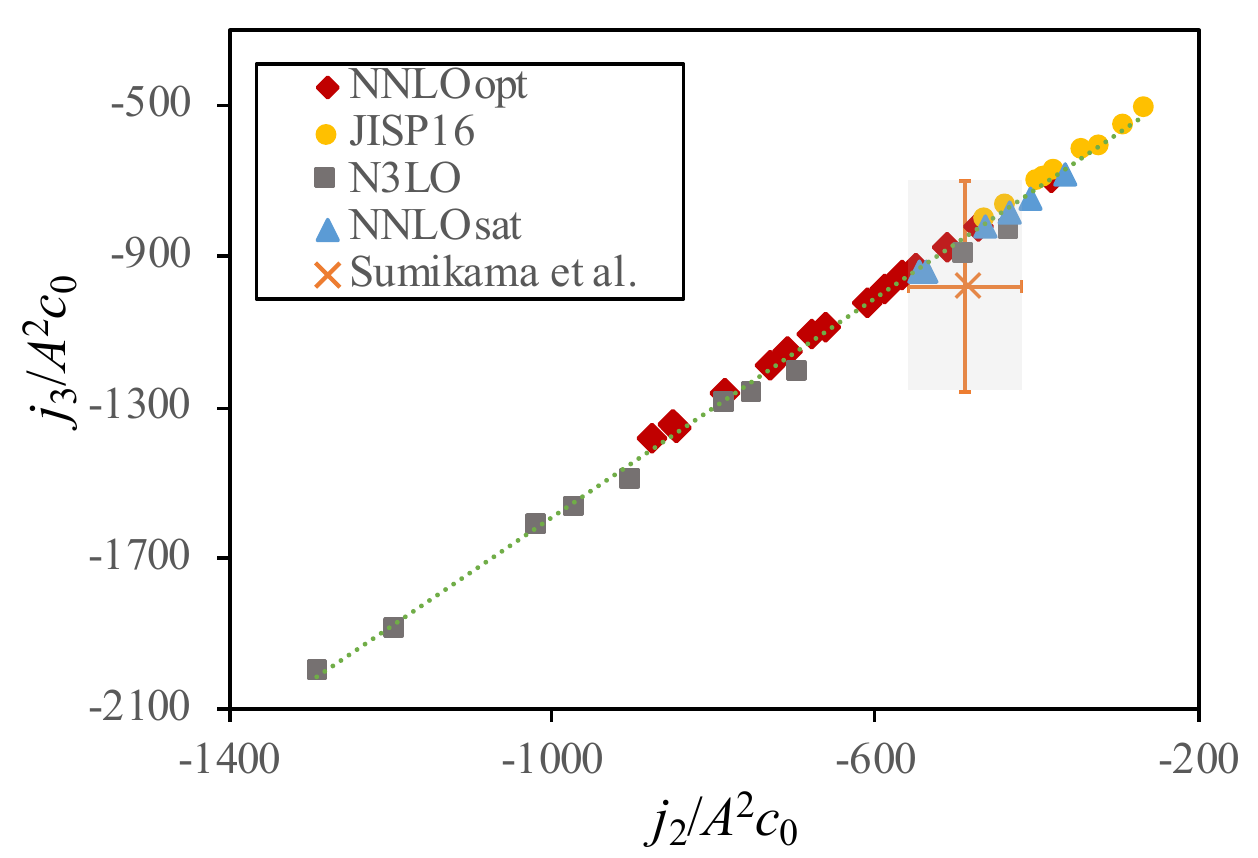}
    \caption{Calculated $j_3/A^2c_0$ vs. $j_2/A^2c_0$  values with \NNLOopt, NNLO$_\mathrm{sat}$, N3LO-EM and JISP16 interactions for the same model spaces as in Fig. \ref{fig:jk_vs_Q}b, along with the values from Sumikama \etal~ \cite{sumikama2011test} (orange cross). The linear fit (dotted green line) is added to guide the eye.
}
    \label{fig:j2_vs_j3}
\end{figure}

Since both $j_2/A^2c_0$ and $j_3/A^2c_0$ for the \Be~ lowest $2^+$ state are strongly correlated with the \Li~ g.s. quadrupole moment (Fig. \ref{fig:jk_vs_Q}), it is expected that there is a strong correlation between $j_2$ and $j_3$. Indeed, the calculated $j_2/A^2c_0$ vs. $j_3/A^2c_0$ for all four interactions under consideration follow an almost perfect linear trendline (Fig. \ref{fig:j2_vs_j3}). It is important to note that the experimentally deduced values from Ref. \cite{sumikama2011test}, which have been determined  over the whole range of the \bt energy spectrum, also agree with the calculated correlation. 

\subsection{Overlaps of \Be~wavefunctions with the  $\alpha+\alpha$ and $^7\mathrm{Li}+$p systems}

\begin{figure*}[ht]
    \centering
   \includegraphics[width=0.99\textwidth]{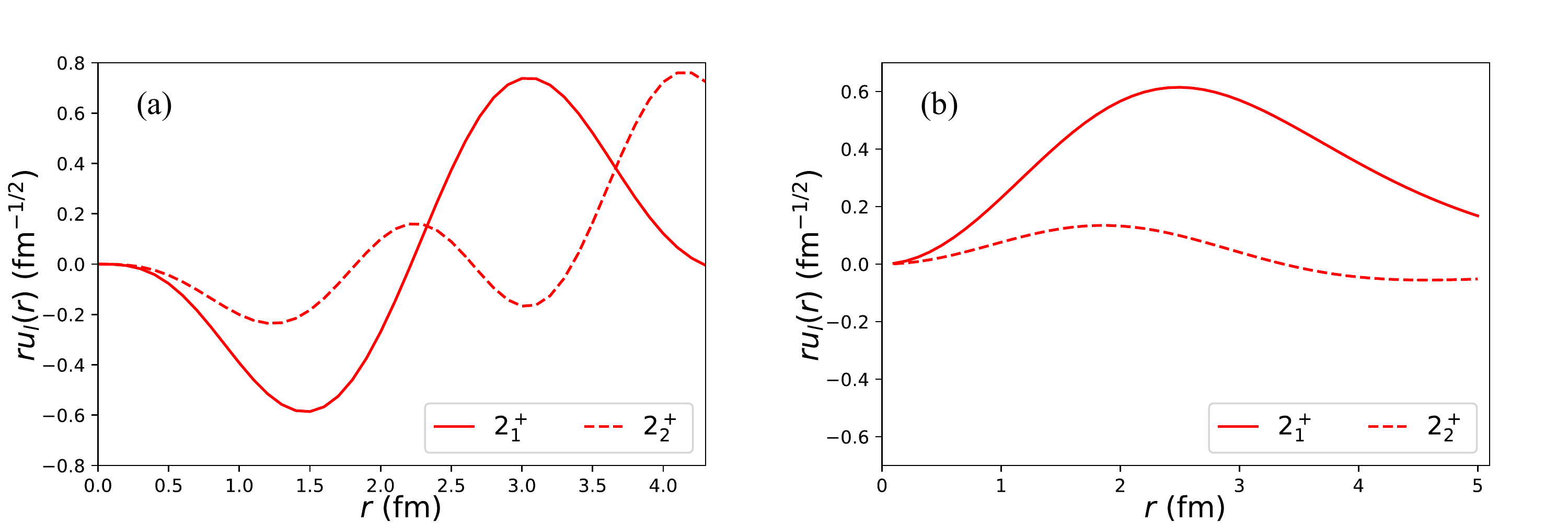}
    \caption{Interior cluster wavefunctions for the lowest two \Be~$2^+$ states with (a) $l=2,\,s=0$  $\braket{^8\mathrm{Be}}{\alpha+\alpha}$ in \Nmax=16 and (b) $l=1,\, s=1$ $\braket{^8\mathrm{Be}}{^7\mathrm{Li}+p}$ in \Nmax=10  (SA-NCSM calculations use  \NNLOopt~and \hw = 15 MeV). 
    }
    \label{fig:a+aJJ4}
\end{figure*}

To study the cluster structure of the lowest $2_1^+$ and $2_2^+$ states, we calculate the overlaps of the many-body \Be~wavefunctions with the $\alpha+\alpha$ and $^7\mathrm{Li}+$p systems, and present the interior $\alpha+\alpha$ cluster wave function (Fig. \ref{fig:a+aJJ4}). The experimental threshold of these systems with respect to the \Be~ground state are $-0.092$ MeV for $\alpha+\alpha$ and $17.255$ MeV for $^7\mathrm{Li}+$p, thus the lowest two $2^+$ states are below the proton separation energy. For the calculations of $\alpha+\alpha$ cluster wavefunctions (Fig. \ref{fig:a+aJJ4}a), we follow the procedure in Ref. \cite{DreyfussLESBDD2020} using the \Be~$2^+$ states and $^4\mathrm{He}$ ground state calculated in \SpR{3} basis. To compose a $J=2$ cluster state with two $\alpha$'s in their ground state and channel spin $s=0$, we consider  the relative orbital momentum $l=2$. Both $2^+$ states have larger peaks at longer distances suggesting alpha clustering, whereas the $2_2^+$ state is more spatially expanded. 
Calculations of the single-nucleon overlaps $\braket{^8\mathrm{Be}}{^7\mathrm{Li}+{\rm p}}$  follow the formalism of Ref. \cite{Navratil04} (Fig. \ref{fig:a+aJJ4}b). Here, the $p$-wave is described by the  $^7\mathrm{Li}$ ground state, which couples with the proton spin to a channel spin $s=1$. We find a large overlap between the $^7\mathrm{Li}+$p and the lowest $2^+$ state, and a comparatively small but non-negligible overlap with the $2_2^+$ state. 

\subsection{\Be~wavefunctions in the symplectic basis}

\begin{figure*}[ht]
    \centering
   \includegraphics[width=0.99\textwidth]{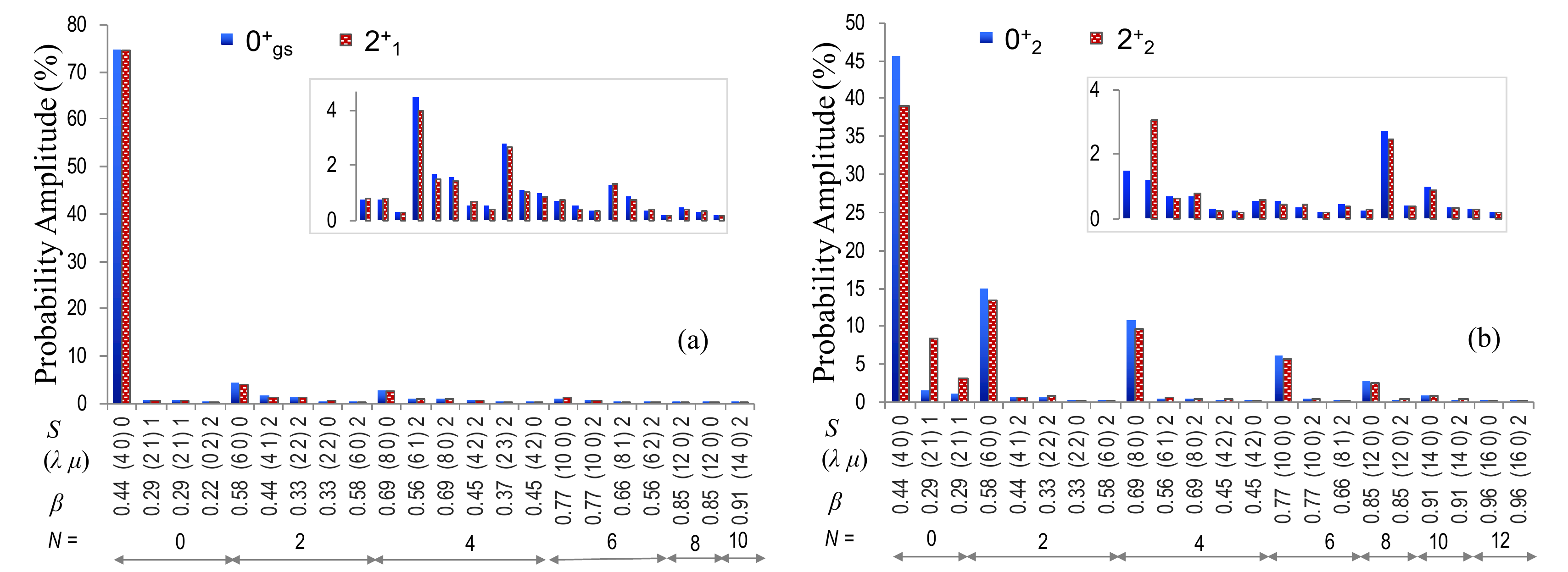}
    \caption{Symplectic \SpR{3} irreps that compose the lowest $0_{\rm g.s.}^+$ and $2_1^+$ states (a), and the second  $0_2^+$ and $2_2^+$ states of \Be; each irrep is specified by its equilibrium shape, labeled by \bt and the corresponding SU(3) labels $(\lambda \, \mu)$ together with total intrinsic spin $S$. $N$ denotes the total HO excitations. Insets: the same irreps but without the terms larger than 5\%.}
    \label{fig:Sp3R}
\end{figure*}

In SA-NCSM the many-body wavefunctions can be decomposed in terms of collective \SpR{3} basis states. As shown in Ref. \cite{DytrychLDRWRBB20}, each irreducible representation (irrep) of the symplectic \SpR{3} group describes an equilibrium shape within a nucleus that vibrates and rotates. These irreps are labeled by the \SU{3} quantum numbers $(\lambda \, \mu)$ of the equilibrium shape with an average deformation given by the familiar shape parameters, the deformation $\beta$ and triaxiality $\gamma$ \cite{CastanosDL88,MustonenGAB2018}.  Typically, the nuclei are comprised  of only a few shapes, often a single shape (a single symplectic irrep) as for the \Be~ground state and the corresponding lowest $2^+$ state (Fig. \ref{fig:Sp3R}a). However, for the $0_2^+$ and $2_2^+$ states, the wavefunction consists of a mixture of three dominant prolate shapes (Fig. \ref{fig:Sp3R}b). Thus, one needs calculations in ultra large model spaces and consideration of all the major shapes to obtain an accurate description of these states. 

\bibliography{recoil.bib,lsu_latest.bib}

\end{document}